\documentstyle[prb,aps,multicol,epsf]{revtex}
\begin{document}
\draft
\title{Charge and magnetic ordering in two-orbital double-exchange
model for manganites}
\author{G. Jackeli$^{a,b}$, N.B. Perkins$^{a}$, and N.M. Plakida$^{a}$}
\address{
$^{a}$Joint Institute for Nuclear Research, Dubna, Moscow region,
141980, Russia \\
$^{b}$Institute of Physics, Georgian Academy of Sciences, Tbilisi, Georgia. }
\maketitle
\begin{abstract}\widetext
Phase diagram of half-doped perovskite manganites  is studied
within the extended  double-exchange model.
To demonstrate the role of orbital degrees of freedom both one- and
two-orbital models are examined.  A rich phase diagram
is obtained in the mean-filed theory at zero temperature as a function of
$J$ (antiferromagnetic (AFM) superexchange interaction) and $V$
(intersite Coulomb repulsion).
For the one-orbital model a charge-ordered (CO) state appears at any value of
$V$ with different types of magnetic order which changes with increasing $J$
from  ferromagnetic (F) to AFM ones of the types A, C and G .
The orbital degeneracy results in appearance of a new
CE-type spin order that is favorable due to  opening of the ``dimerization''
gap  at the Fermi surface. In addition, the CO state appears
only for $V > V_{c}$ for F and CE states while  C-type AFM state 
disappears
and A-type AFM state is observed only at small values of $V$
as a charge disordered one.
The relevance of our results to the experimental data are  discussed.
\end{abstract}
\pacs{PACS numbers: 75.25.+z, 75.30.Et, 75.30.Vn, 71.45.Lr}
\begin{multicols}{2}
\narrowtext
\newpage
\section{Introduction}
Since early fifties \cite{wol} the physics of manganites
challenge our current understanding of transition-metal
oxides, and define both theoretical and experimental
research problem that involves charge, spin, lattice  and orbital degrees
of freedom.
Recently in a modern systematic experimental studies
a very rich phase diagram (see, for example, Ref.\onlinecite{diagr}) depending
on the doping concentration, temperature and pressure was obtained
in the  doped manganese oxides with perovskite structure
${\rm R}_{1-x}{\rm B}_x{\rm MnO}_3$ (where
${\rm R}$ is trivalent rare-earth and ${\rm B}$ is divalent alkaline ion,
respectively).
At different doping concentration
a full variety of magnetically ordered states  such as
antiferromagnetic (AFM) insulator, ferromagnetic (FM)
metal and  charge ordered (CO) insulator were  observed.
Many efforts
have been made by theoreticians to understand
it based on various models
and approaches. Historically, double exchange (DE) model\cite{DE}
was the basic one.
In this model $t_{2g}$-electrons are localized, whereas the $e_g$-electrons
are mobile and use ${\rm O}$ $p$-orbitals as a bridge between ${\rm Mn}$ ions.
The hopping of itinerant electrons   together with a very strong on-site
Hund's coupling drives core spins to align parallel.
Qualitatively DE model gave appropriate interpretation of the  phase diagram
at the doping range $0.2<x<0.5$ where FM metallic behavior was observed.

One of the subtle aspects of the perovskite manganites is the charge ordered
state observed in almost all such  compounds
at half-doping.\cite{rad,cheong,tom,kuw}
A direct evidence of the CO
state in half-doped manganites has been provided by the electron
diffraction for ${\rm La}_{0.5}{\rm Ca}_{0.5}{\rm MnO}_3$.\cite{cheong}
Similar observations have also been reported for
${\rm Pr}_{0.5}{\rm Sr}_{0.5}{\rm MnO}_3$
\cite{tom} and in ${\rm Nd}_{0.5}{\rm Sr}_{0.5}{\rm MnO}_3$.\cite{kuw}
CO state is characterized by an alternating ${\rm Mn}^{3+}$ and ${\rm Mn}^{4}$
ions arrangement in $x-y$
plane with the charge stacking in $z$-direction. In CO state
these systems show an insulating behavior with a
very peculiar form of AFM spin
ordering. The observed magnetic structure is a  CE-type
and consists of quasi one-dimensional ferromagnetic zig-zag chains coupled
antiferromagnetically. In addition, these systems show
$d_{3x^2-r^2}$/$d_{3y^2-r^2}$ orbital ordering.
Another noteworthy observations were done by studying
${\rm Pr}_{1-x}({\rm Ca}_{1-y}{\rm Sr}_{y})_{x}{\rm MnO}_3$ crystals with
controlled one-electron bandwidth. As already mentioned above at half-doping
${\rm Pr}_{0.5}{\rm Ca}_{0.5}{\rm MnO}_3$ has a CO CE-type insulating state.
However, by  substitution ${\rm Ca}$ with ${\rm Sr}$ leading to the increase
of the carrier bandwidth, one induces the collapse of the CO insulating state,
and the A-type metallic state with $d_{x^2-y^2}$ orbital ordering
is realized in
${\rm Pr}_{0.5}{\rm Sr}_{0.5}{\rm MnO}_3$.\cite{kaw}
The coexistence of the A-type spin ordered and CE-type spin/charge  ordered
states has been detected in the bilayer
${\rm La}{\rm Sr}_{2}{\rm Mn}_2{\rm O}_7$
\cite{kub} and  three-dimensional     
${\rm Nd}_{0.5}{\rm Sr_{0.5}}{\rm MnO}_3$.\cite{kuw}
These results indicate the competition between the metallic A-type $d_{x^2-y^2}$
orbital ordering and the insulating CE-type $d_{3x^2-r^2}/d_{3y^2-r^2}$
orbital ordering at half-doping and
demonstrate the importance of the magnetic, charge and
orbital order coupling in these compounds.

In recent publications
it was shown that the DE anisotropy resulted from the orbital degeneracy with
  the peculiar $e_g$ transfer amplitudes is important and
to be a key point in explaining the
  various types of AFM ordering.\cite{gor,brink}
Until now most of the theoretical studies of CO state were done
  in the framework  of the one-orbital model
ignoring the double  degeneracy of $e_g$ orbitals.\cite{RPA,min,yu}
The detailed mean-filed analysis of phase diagram of one-orbital DE model in
the presence of both on-site and inter-site Coulomb terms has been given in Ref.\onlinecite{sat}. It has been shown, that in the vicinity of half-doping the double-exchange gain of energy is considerably suppressed by the inter-site Coulomb interaction that favors charge-ordered state.
Recently, the CO state within the   two-orbital model has been investigated by
  the projection perturbation techniques combined with the coherent state
  formalism and by Monte Carlo simulations in Refs.~\onlinecite{shen} and
  \onlinecite{yuno}, respectively.
In Ref.\onlinecite{shen} the origin  of CO has been attributed to the effective
  particle-hole interaction and CO state with C-type  of the spin ordering at
  $x=0.5$ has been obtained. The Authors of Ref.\onlinecite{yuno}
have shown that the
  experimentally observed charge, spin, and orbital ordering could be
stabilized due to  Jahn-Teller phonons.

In the present paper we investigate the role of the orbital
degeneracy in the CO state based on  the two-orbital DE model including the
intersite Coulomb interaction. We adopt the mean-field (MF)  approximation to
derive the ground state phase diagram in the two-orbital model,
and compare it to  that of the corresponding one-orbital model.
We argue that the orbital degeneracy
together with the peculiar $e_g$ transfer amplitude  has a drastic effect on
the phase diagram and is important in obtaining the realistic
magnetic/charge/orbital ordering observed in half-doped manganites.
The paper is organized as follows:
In the next section the model Hamiltonian is presented and the mean-field scheme
is formulated. The ground state phase diagrams of the one- and two-orbital models
are derived and compared in Sec.III. Sec.IV summarizes  our main results.
In the Appendix the canonical transformation diagonalizing the
MF Hamiltonian and the resulted  band structure of various magnetic phases
are presented.

\section{Model and Formulation}
We start with the two orbital ferromagnetic Kondo lattice model
supplemented by the intersite Coulomb repulsion
\begin{eqnarray}
H&=&-\sum_{\langle ij\rangle,\sigma}
t_{ij}^{\alpha\beta}
\left[d_{i\sigma\alpha }^{\dagger}d_{j\sigma\beta}+H.c.\right]-
J_{\rm H}\sum_{i}{\bf S}_i{\bf \sigma}_i \nonumber\\
&+&J\sum_{\langle ij\rangle}{\bf S}_{i}{\bf S}_{j}
+V\sum_{\langle ij\rangle}n_{i}n_{j}-\mu\sum_{i}n_{i}.
\label{1}
\end{eqnarray}
The first term  of Eq.(\ref{1}) describes an electron
hopping between the two $e_{g}$ orbitals of the  nearest neighbor (NN) Mn-ions.
The orbitals $d_{3z^{2}-r^{2}}$ and $d_{x^{2}-y^{2}}$ correspond to
$\alpha(\beta)=1$ and 2, respectively.
Due to the shape of the $e_g$
orbitals, their  hybridization is different in
the three cubic directions that leads to direction dependent hopping
with the anisotropic transfer matrix elements  $t_{ij}^{\alpha \beta}$
given by
\begin{eqnarray}
t_{x/y}^{\alpha \beta}=
t\left(\!\!\begin{array}{cc}
1/4 &\!
\mp\sqrt{3}/4\\
\mp\sqrt{3}/4&\!3/4
\end{array}
\!\!\right)~,\;\;\;
t_{z}^{\alpha \beta}=
t\left(\!\!\begin{array}{cc}
1 &\!\;\;
0\\
0&\!\;\;0
\end{array}
\!\!\right)\;\;\;  .
\label{2}
\end{eqnarray}
The second term in Eq.(\ref{1}) describes the Hund's coupling between the spins of
localized $t_{2g}$-~electrons ${\bf S}_i$  and the itinerant $e_{g}$
electrons with spin ${\bf  \sigma}_{i}$.
The  superexchange (SE) interaction  of localized
spins between the NN sites is given by $J$, $V$ represents
the  inter-site Coulomb repulsion of $e_{g}$
electrons, $n_{i}$  is the particle number operator
and $\mu$ is the chemical potential.
The effect of the on-site Coulomb interaction that is not included in our model
Hamiltonian will be discussed later.

We study the Hamiltonian (\ref{1}) within the MF approximation,
which is set up by  introducing the order parameter for static
charge-density wave  of the form
$\langle n_{i}\rangle=n+\delta n \exp(i{\bf Q}{\bf R}_{i})$, with $n$
being the electron density and ${\bf Q}=(\pi,\pi,\pi)$.
Further, we treat localized spin subsystem classically and
assume a  strong Hund's  coupling $J_H\gg zt/S$.
In this limit one may  take the local spin quantization axis parallel
to  $t_{2g}$-spins and in the rotated bases
retain only "spin-up" components of the mobile electrons.
Then the transfer integral between the NN sites is modified
through relative angle of the
$t_{2g}$-spins at the $i$ and $j$ sites as
$\tilde{t}_{ij}^{\alpha \beta}=
t_{ij}^{\alpha \beta}\cos(\theta_{ij}/2)$, where $\theta_{ij}$
is the relative angle of the $t_{2g}$-spins.
We consider the following  magnetic phases that competes:
i) Ferromagnetic configuration (F-type spin ordering) with
$\theta_{xy}=\theta_{z}=0$ ( $\theta_{xy}$ and $\theta_{z}$ are the angels
between the neighboring spin in $xy$-plane and  $z$-direction, respectively,
ii) Layer-type  antiferromagnetic configuration
(A-type spin ordering) -- the local spins are parallel
in the planes and antiferromagnetically aligned between the neighboring planes,
that corresponds  to $\theta_{xy}=0$ and $\theta_{z}=\pi$.
iii) Chain-type antiferromagnetic configuration
(C-type spin ordering) -- the local spins are parallel in the
straight chains and
antiferromagnetically coupled between the chains --
$\theta_{xy}=\pi$  and  $\theta_{z}=0$.
iv) Neel-type antiferromagnetic configuration (G-type spin ordering)
with all spins being antiparallel --- $\theta_{xy}=\theta_{z}=\pi$.
v) CE-type spin ordering with zig-zag ferromagnetic chains coupled antiferromagnetically.

As a result we come to the following MF Hamiltonian:
\begin{eqnarray}
H&=&-\sum_{\langle ij\rangle,\alpha,\beta}
\tilde{t}_{ij}^{\alpha\beta}
\left[d_{i\alpha }^{\dagger}d_{j\beta}+H.c.\right]-
\Delta\sum_{ i}e^{i{\bf Q}{\bf R}_{i}}n_{i} \nonumber\\
&-&\mu\sum_{i}n_{i}
+2(d-3)JS^2 N~,
\label{3}
\end{eqnarray}
where $\Delta=zV\delta n$,  $z=6$ for 3-dimensional cubic lattice,
and $d$ is dimensionality of the magnetic order ( $d=0,1,2,$ and $3$, respectively
for  G-, C-, A-, and F- type spin ordering).
In Eq.(\ref{3}) the zero of the energy is chosen in such a manner that
the SE energy  vanishes in the FM state.
To obtain phase diagram we need to compare the free energies
of all possible magnetic configurations.

\section{Phase diagram}
\subsection{One-orbital model}
In order to incorporate the role of orbital
degeneracy,  first  we consider the one orbital model
ignoring the double degeneracy of $e_{g}$ orbitals.
Retaining only the one orbital per Mn-ion and assuming the isotropic transfer amplitude,  the electronic part of the MF Hamiltonian in ${\bf k}$-space is written as:
\begin{eqnarray}
H^{\rm 1orb}_{\rm el}=-\sum _{\bf k} (\tilde{t}_{\bf k}+\mu) d_{\bf k}^{\dagger}d_{\bf k}
-\Delta \sum _{\bf k} d_{\bf k}^{\dagger} d_{{\bf k}+{\bf Q}}~,
\label{4}
\end{eqnarray}
with
\begin{equation}
\tilde{t}_{\bf k}=2t\sum_{i=1}^{d}\cos {\bf k}_{i}~, \;\;\;\;
({ k}_{1},{ k}_{2},{ k}_{3})=(k_{x},k_{y},k_{z}).
\label{5}
\end{equation}
The above Hamiltonian (\ref{4})
is easily diagonalized by the following canonical
transformation:
\begin{equation}
d^{}_{{\bf k}}=u_{{\bf k}}c^{}_{1,{\bf k}}+v_{{\bf k}}c^{}_{2,{\bf k}},\;\;
                 d^{}_{{\bf k}+{\bf Q}}=-v_{{\bf k}}c^{}_{1,{\bf k}}+u_{{\bf k}}c^{}_{2,{\bf k}}
\label{6}
\end{equation}
with
\begin{eqnarray}
u_{{\bf k}}&=&\frac{1}{\sqrt{2}}\left[
1-\frac{\tilde {t}_{\bf k}}{\varepsilon_{\bf k}}
  \right]^{\frac{1}{2}},\;
v_{{\bf k}}=\frac{1}{\sqrt{2}}\left[
1+\frac{\tilde {t}_{\bf k}}{\varepsilon_{\bf k}}
  \right]^{\frac{1}{2}} , \nonumber\\
\varepsilon_{{\bf k}}&=&\sqrt{\tilde{t}_{\bf k}^2+\Delta^2} \; .
\label{7}
\end{eqnarray}
In terms of the $c$-operators, the one particle Hamiltonian reads
\begin{equation}
H = \sum\limits_{{\bf k},\alpha}^{} (-1)^{\alpha} \varepsilon({\bf k})
c^{\dag}_{{\bf k},\alpha}c^{}_{{\bf k},\alpha} \ , \  \alpha =1,2\ .
\label{8}
\end{equation}
At half-filling the chemical potential lies inside the gap ($\mu=0$) and recalling that
$\Delta=zV\delta n$ we receive a self-consistent equation for
the order parameter
\begin{equation}
1=\frac{zV}{2N}\sum_{{\bf k}} \frac{\tanh \beta
\varepsilon_{\bf k} /2}{\varepsilon_{\bf k}}~.
\label{9}
\end{equation}

\begin{figure}
      \epsfysize=55mm
      \centerline{\epsffile{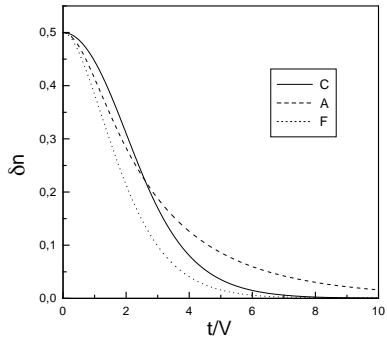}}
\caption{The zero temperature charge order parameter $\delta n$ as a function
of t/V for the one-orbital model and for a different
(F-, A-, and C-type) spin ordered states. Here $t$ is the hopping amplitude and $V$ is the intersite Coulomb repulsion.}
\label{f1}
\end{figure}
In Fig.\ref{f1} the overall behavior of the order parameter $\delta n$
as a function of $t/V$ is presented for various magnetic configuration.
Since the wave vector summation in the right hand side of Eq.(\ref{9})
diverges in the limit $\Delta\rightarrow 0$ there exist a nontrivial
solution even at $V\rightarrow 0$ and hence a transition from
homogeneous to CO state is continuous. We also note that $\delta n$ diminishes
exponentially with increasing the bandwidth  (see Fig.\ref{f1}) indicating that the
transition between the  homogeneous and the CO state is a result of the
competition
between the kinetic and the electrostatic energy.

\begin{figure}
      \epsfysize=55mm
      \centerline{\epsffile{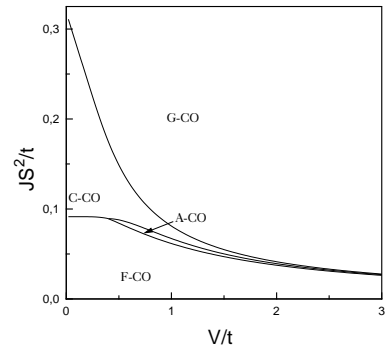}}
\caption{Phase digram of the one-orbital model in $JS^2/t$ and $V/t$ parameter
plane. Here A, C, and G denotes the A-, C-, and G-type antiferromagnetic ordering, respectively, F denotes the ferromagnetic phase and CO is charge ordering.
The solid lines stand for the first--order  phase boundaries.}
\label{f2}
\end{figure}

By comparing the free energies of different magnetic configuration we obtain
the phase diagram as shown in Fig.\ref{f2}. At small  $V$, with
increasing $J$ the system, starting from the F-CO phase, first enters to
the C-CO phase and then to the G-CO state. Since  the gain in the
magnetic energy when the system moves
from A- to C-phase is larger then the gain in the kinetic energy
in C to A transition the A-CO phase is absent in this part of Phase diagram.
With increasing of $V$ at $V\simeq 0.5t$
the CO gap in C-CO phase overcomes  that one in A-CO phase
that results in opening of small window of A-CO phase  in the phase digram.
We also note that with increasing of $V$ the SE coupling
needed to stabilize the AFM configuration decreases since the
bandwidth effect is overshadowed when the gap becomes larger.
The different magnetic structures in Fig.\ref{f2} are separated by the first-order boundaries. There is a jump in the charge order parameter across the phase boundaries, since the value of order parameter $\delta n$ depends
on the effective bandwidth and hence on the underlying magnetic structure,
as can be seen from Fig. 2.

\subsection{Two orbital model}
To describe the effect of orbital degeneracy, we consider the MF Hamiltonian
(\ref{3}) with anisotropic hopping amplitude. In the
momentum space the electronic part of the Hamiltonian reads as:
\begin{eqnarray}
H_{\rm el}^{\rm 2orb}&=&\sum_{{\bf k},\alpha,\beta}
[\varepsilon_{{\bf k}}^{\alpha\beta}-\mu\delta_{\alpha\beta}]
d_{{\bf k}\alpha }^{\dagger}d_{{\bf k}\beta}-
\Delta\sum_{ {\bf k},\alpha}d_{{\bf k},\alpha}^{\dagger} d_{{\bf k}+{\bf Q}
,\alpha}
\label{10}
\end{eqnarray}
with
\begin{eqnarray}
\varepsilon_{{\bf k}}^{11}&=&-\frac{1}{2}
{\tilde t}_{xy}(\cos {\bf k}_{x}+\cos {\bf k}_{y})
-2{\tilde t}_{z}\cos {\bf k}_{z}, \nonumber \\
\varepsilon_{{\bf k}}^{12}&=&\varepsilon_{{\bf k}}^{21}
=-\frac{\sqrt{3}}{2}{\tilde t}_{xy}(\cos {\bf k}_{x}-\cos {\bf k}_{y}),\nonumber \\
\varepsilon_{{\bf k}}^{22}&=&-\frac{3}{2}t_{xy}(\cos k_{x}+\cos k_{y})
\label{11}
\end{eqnarray}
and ${\tilde t}_{xy} = t\cos \theta_{xy}, {\tilde t}_{z}=t\cos \theta_z$.
\begin{figure}
      \epsfysize=55mm
      \centerline{\epsffile{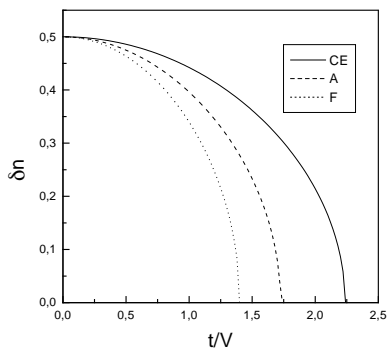}}
\caption{Order parameter $\delta n$ versus t/V for the two-orbital
model and for a different
(F-, A-, and CE-type) spin ordered states.}
\label{f3}
\end{figure}

The diagonalization of this Hamiltonian ( see Appendix) leads to the four
band model.
In the case of  F- and A-type spin ordering and
at the filling corresponding to one electron per two Mn-ions
(half-doped case) the gap is not opened at the Fermi surface, and
the chemical potential moves down with the lower two bands.
We solve  the gap equation
self-consistently with one for the chemical potential.
As it seen in Fig.\ref{f3},
the transition to the charge ordered state is not continuous and there exists
a critical value $V_{\rm c}$  above which the ordered state
is favorable
($V_{\rm c}^{\rm F}\simeq 0.72 t$ and
$V_{\rm c}^{\rm A}\simeq 0.58t$ for F- and A-type spin ordering, respectively).
As for the $C$-type spin ordering,  there is no difference between
the one and two orbital models in this sense.
The existence of the additional orbital only introduce the localized level
which is empty at the filling we consider.

\begin{figure}
\epsfysize=40mm
      \centerline{\epsffile{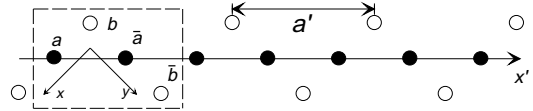}}
\caption{The one isolated  zig-zag shaped ferromagnetic chain. The real CE-type spin structure is given  by this quasi-one-dimensional ferromagnets coupled antiferromagneticaly. White and black circles denote  ${\rm Mn}^{4+}$ and
${\rm Mn}^{3+}$ ions respectively.
The dashed line shows the unit cell and $a'$ is the
lattice parameter along the zig-zag}
\label{f4}
\end{figure}

The above Hamiltonian (10) describes
the F-, A-, C-, and G-type spin ordering on an equal footing.
However the presence of additional orbital degree of freedom, with
the peculiar anisotropic transfer amplitudes $t_{x,y,z}^{\alpha\beta}$
[see Eq.(\ref{2})] results in the anisotropic DE interaction
\cite{sol} and
may lead to the stabilization  of CE spin ordering.
Let us consider one  zig-zag  with two ferromagnetic bonds alternated
in $x$ and $y$ directions (Fig.\ref{f4}). The corner and middle
sites are denoted by $a (\bar a)$ and $b (\bar b)$, respectively,
and the unit cell is given by four nonequivalent atoms.
\mbox{}\\
For further discussions it is  convenient to adopt the following bases of the $e_{g}$
orbitals at nonequivalent sites:
\cite{sol}
$|1\rangle=d_{3x^2-r^2}$, $|2\rangle=d_{y^2-z^2}$, and
$|\bar{1}\rangle=d_{3y^2-r^2}$
$|{\bar 2}\rangle=d_{x^2-z^2}$ on $a$, $b$ and ${\bar a}$, ${\bar b}$
sites, respectively [see, Fig.\ref{f4}]. In the new  bases the transfer
matrix elements are  given by
\begin{eqnarray}
t_{1}^{\alpha \beta}=
t\left(\!\!\begin{array}{cc}
1 &\!0
\\0&\!0
\end{array}
\!\!\right)~,\;\;\;
t_{2}^{\alpha \beta}=
t\left(\!\!\begin{array}{cc}
-1/2 &\!\;\;
0\\
\sqrt{3}/2&\!\;\;0
\end{array}
\!\!\right)\;\;\;
\label{12}
\end{eqnarray}
between $a$--$b$ ( ${\bar a}$--${\bar b}$) and  $b$--${\bar a}$
( ${\bar b}$--$a$) NN sites, respectively.
As a result the zig-zag chain  is modeled as a dimerized one with the
alternating hopping amplitude
and can be described by the following  Hamiltonian
\end{multicols}
\widetext
\begin{eqnarray}
H^{\rm CE}_{\rm el}
=&-&\sum_{i,\alpha,\beta}\left\{\left[
t_{1}^{\alpha\beta}
\{a_{i\alpha }^{\dagger}b_{i\beta}+{\bar a}_{i\alpha }^{\dagger}
{\bar b}_{i\beta}\}\right.
+\left.t_{2}^{\alpha\beta}
\{b_{i\alpha }^{\dagger}{\bar a}_{i\beta}+{\bar b}_{i\alpha }^{\dagger}
a_{i+1\beta}\}+H.c.\right]
\right.\nonumber\\
&-&(\mu+\Delta)
\{a_{i\alpha }^{\dagger}a_{i\alpha}+
{\bar a}_{i\alpha }^{\dagger}{\bar a}_{i\alpha}\}
-(\mu-\Delta)
\left.\{b_{i\alpha }^{\dagger}b_{i\alpha}+
{\bar b}_{i\alpha }^{\dagger}{\bar b}_{i\alpha}\}\right\}
\label{13}
\end{eqnarray}
\begin{multicols}{2}
\narrowtext
where $i$ runs along the zig-zag and denotes the number of unit cell.
Diagonalization of the above Hamiltonian ( see Appendix) leads to the
complicated band structure consisting of bonding and antibonding bands,
$E_{a,b}=\pm\sqrt{\Delta^2+t^2(2-\cos(k/2))}$, and nonbonding
states $E_{\pm}=\pm\Delta$.
Due to the topology of the zig-zag structure, only the directional
$d_{3x^2-r^2} (d_{3y^2-r^2})$ orbitals at
$a({\bar a})$ sites give the input in the low energy bonding state 
[see Appendix for a details].
While for the  $b({\bar b})$ sites both two orthogonal orbitals do
contribute.
Therefore, the orbital degeneracy is removed on the middle site sublattice and the carriers on this sublattice will occupy the directional orbitals, leading to the polarized orbital state with $d_{3x^2-r^2}/d_{3y^2-r^2}$ orbital ordering. 
We also emphasize, that at half-doping the
bonding band is full and the system is a band insulator even in the absence of
the charge ordering. The onset of the charge ordering renormalizes the gap to
higher value.
The behavior of the charge order parameter is depicted in Fig.3. The transition to charge ordered state takes place at
$V_{\rm c}^{\rm CE}\simeq 0.44t$, that is lower then that one for A- and F- type spin ordering. A smaller value of intersite Coulomb repulsion is needed to introduce the charge ordered state in the state with the lower dimension of ferromagnetic
component.
\begin{figure}
\epsfysize=55mm
      \centerline{\epsffile{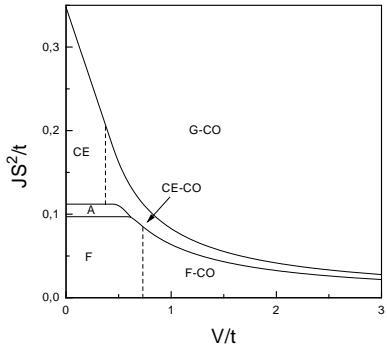}}
\caption{The phase diagram of the two-orbital model ($JS^2/t$ vs $V/t4$)
. The dashed lines stand for the second-order boundaries and 
separate the uniform state from the
charge ordered one in the corresponding magnetic phases.}
\label{f5}
\end{figure}

The ground state phase diagram of the two-orbital model is given in Fig.5.
The phases with different magnetic structures are separated by the first-order boundaries (soled lines in Fig. 5). The transition from the charge--disordered to the charge--ordered version of the given magnetic structure is continuous,
there is no jump  in the  charge order parameter during the transition which does not change the symmetry of underlying magnetic structure ( the corresponding second-order boundaries are denoted by the dashed lines in Fig.5).
At small value of $V$ with increasing of SE coupling the system goes through
four magnetic phases F, A, CE, and G consequently.
Due to the
additional orbital degree of freedom, unlike to the one orbital model, there exist a finite phase space of A-type spin ordering in the small $V$ region $V<V_{\rm c}^{A}$.
However, for $V>V_{\rm c}^{A}$, when the charge ordering is introduced in the A
phase, and hence the bandwidth effect  is suppressed, CE-CO phase wins and
the A-CO phase is never realized in the phase diagram of the present model.

\section{Conclusion}
In summary, we have investigated  the ground state phase diagram  of the
one- and two-orbital extended double exchange  models
within the mean filed approximation at half-doping.
In the case of the one-orbital model
the MF theory predicts the continuous phase transition to the
charge ordered state due to the intersite  Coulomb interaction
$V$.
In the two-orbital model the character of the transition
is changed by the  introducing  the nonzero critical value of $V$.
While the transition to charge ordered state with given magnetic structure is continuous, there is jump in of charge order parameter across 
the phase boundaries
between the states with different symmetry of magnetic structure.
Depending on the intersite Coulomb interaction $V$ and
and superexchange coupling $J$
different types of spin ordering
(F-, A-, G-type) accompanied by the charge ordering may take place in
the ground state of the one-orbital model.
The presence of orbital degeneracy with the peculiar anisotropic $e_g$
transfer amplitude introduces the new magnetic state, CE-type
spin ordering, in the phase diagram of the two-orbital model.
The C-type spin ordering is never achieved within this model due to its
instability against the effective "dimerization" and formation of the zig-zag
ferromagnetic order.
The alternation of the ferromagnetic bonds in $x/y$ directions leads to the
alternation of the hopping amplitude. As a result the bare band is splited
into bonding and antibonding states and the "dimerization" gap opens on the
Fermi surface  at half-doping.
The CE-type spin ordered state is accompanied with
$d_{3x^2-r^2}/d_{3y^2-r^2}$ orbital ordering, originated from the topology of
the zig-zag structure.
The CE-type spin/charge  state also wins  the CO state with A-type
AFM ordering and there exists  the phase boundary between the 
CE spin/charge  ordered
state and charge disordered (CD) A-type state  [see Fig.5]. 
The experimentally
detected   competition between A-CD  and
CE-CO states in the half-doped manganites could indicate that the
 parameters of the system are close to this phase boundary.
 Therefore the small change of the bandwidth may drive the
system from one to other state. That is also suggested by
the huge oxygen isotope effect observed in La-Pr compounds. \cite{bab,bal}
A substitution of $O^{16}$ by the isotope  $O^{18}$ narrows the carrier bandwidth due
to the polaronic effect and as a result the electrostatic energy
might  overcome the kinetic energy and the charge ordered insulating
state might be  established.

We would like to point out that there exist other physical 
factors and ingredients not included in present treatment that 
may stabilize the particular state and modify the phase diagram 
(quantum nature of the spins, coupling to the Jan-Teller phonons,  
and on site Coulomb interaction).  As it was recently discussed in 
Ref.\onlinecite{brink1} the interorbital on-site Coulomb interaction 
$U^{\prime}$ could be
responsible for the experimentally observed charge ordered state.
In the CE structure the orbital degeneracy is removed on the sublattice
composed by the middle sites and the low energy state corresponds
to the directional orbitals (see Appendix of the present paper). 
If charge carriers  occupy the corner 
sites   there will be  a positive
contribution from the $U^{\prime}$ term to the system energy.
While, if only 
 middle site sublattice is occupied this positive contribution
disappears since the onsite Coulomb term has the zero matrix element in this 
case. Therefore, $U^{\prime}$ term in CE-type spin ordered state acts
as a source of the charge ordering, but in a rather deferent way then the
intersite Coulomb repulsion. 
Since the middle sites are stacked in
$z$-direction $U^{\prime}$ induces experimentally observed 
$(\pi,\pi,0)$ charge ordering instead of
the Wigner crystal type CO  favored by the intersite Coulomb interaction
considered here.
Nevertheless, we believe that the main effect of the orbital 
degeneracy on the interplay of different type of magnetic ordering 
in charge-ordered state is at least qualitatively captured by the 
present treatment indicating the crucial importance of the  electronic 
state degeneracy on the phase diagram.

\acknowledgments
Financial support by the
the INTAS Program, Grants No 97-0963 and No 97-11066, are  acknowledged.

\appendix
\section*{Diagonalization in momentum space}
\subsection{Two orbital model}
Diagonalization of the Hamiltonian (\ref{10}) can be done by two
subsequent canonical transformation. First, we diagonalize the
free part of the Hamiltonian by introducing the new fermionic operators
\begin{equation}
 \left\{
 \begin{array}{l}
d_{{\bf k}1}={\tilde u}_{{\bf k}}{\bar d}_{{\bf k}1}-{\tilde v}_{{\bf k}}{\bar d_{{\bf k}2}},\;\;
d_{{\bf k}2}=-{\tilde v}_{{\bf k}}{\bar d_{{\bf k}1}}+{\tilde u}_{{\bf k}}{\bar d_{{\bf k}2}}
\end{array} \right.
\label{A1}
\end{equation}
with
\begin{eqnarray}
\left\{ \begin{array}{l}
         {\displaystyle  {\tilde u}_{{\bf k}} =\frac{|\varepsilon^{12}_{{\bf k}}|}{\sqrt{
(\varepsilon_{{\bf k},1}-\varepsilon_{{\bf k},2})(\varepsilon_{{\bf k},1}-\varepsilon^{11}_{{\bf k}})}}} \\ \\
                 {\displaystyle {\tilde v}_{{\bf k}} =\frac{\varepsilon^{12}_{{\bf k}}}{\sqrt{
(\varepsilon_{{\bf k},1}-\varepsilon_{{\bf k},2})(\varepsilon_{{\bf k},1}-\varepsilon^{22}_{{\bf k}})}}}  \\ \\
                  {\displaystyle \varepsilon_{{\bf k},1/2}=\frac{1}{2}
\left\{\varepsilon^{11}_{{\bf k}}+\varepsilon^{22}_{{\bf k}}\pm
\sqrt{(\varepsilon^{11}_{\bf k}-\varepsilon^{22}_{\bf k})^2+
4(\varepsilon^{12}_{{\bf k}})^2}\right\}
. }
\end{array}
\right.
\label{A2}
\end{eqnarray}
One finds the effective Hamiltonian of the form
\begin{eqnarray}
H^{\rm 2orb}_{\rm el}& =& \sum\limits_{{\bf k},i}
\varepsilon_{{\bf k},i}d^{\dagger}_{{\bf k}i}d_{{\bf k}i}
-\delta\sum_{\bf k}{\text sgn}(\varepsilon^{12}_{\bf k})\nonumber\\
&\times&\left[d^{\dagger}_{{\bf k},1}d_{{\bf k}+{\bf Q}2}-
d^{\dagger}_{{\bf k}2}d_{{\bf k}+{\bf Q}1}\right].
\label{A3}
\end{eqnarray}
Further we perform the transformation similar to (\ref{6}) and introduce
the new sets of fermionic operators as
\begin{eqnarray*}
a^{}_{{\bf k}1}=
{\bar u}_{{\bf k}1}{\bar d}^{}_{{\bf k}1}+
{\bar v}_{{\bf k}1}{\bar d}^{}_{{\bf k}+{\bf Q}2},\;
a^{}_{{\bf k}2}={\bar u}_{{\bf k}1}{\bar d}^{}_{{\bf k}+{\bf Q}2}
-{\bar v}_{{\bf k}1}{\bar d}^{}_{{\bf k}1},
\nonumber\\
a^{}_{{\bf k}3}=
{\bar u}_{{\bf k}2}{\bar d}^{}_{{\bf k}2}+
{\bar v}_{{\bf k}2}{\bar d}^{}_{{\bf k}+{\bf Q}1},\;
a^{}_{{\bf k}4}={\bar u}_{{\bf k}2}{\bar d}^{}_{{\bf k}+{\bf Q}1}
-{\bar v}_{{\bf k}2}{\bar d}^{}_{{\bf k}2},
\end{eqnarray*}
where
\begin{eqnarray*}
{\bar u}_{{\bf k}i}&=&\frac{1}{\sqrt{2}}\left[
1-\frac{\varepsilon_{{\bf k}i}}{E_{{\bf k}i}}
  \right]^{\frac{1}{2}}\!\!\!,\;\;\;
                 {\bar  v}_{{\bf k}i}=(-1)^{i}
\frac{\text{sgn}(\varepsilon^{12}_{\bf k})}{\sqrt{2}}\left[
1+\frac{\varepsilon _{{\bf k}i}}{E_{{\bf k}i}}
  \right]^{\frac{1}{2}}\nonumber  \\
E_{{\bf k}i}&=&\sqrt{\varepsilon _{{\bf k}i}^2+\Delta^2},
\hspace*{2cm} i=1,2.
\end{eqnarray*}
As a result we get the following four band Hamiltonian
\begin{eqnarray}
H^{\rm 2orb}_{\rm el}&=&\sum\limits_{{\bf k}}\left\{
E_{{\bf k}1}[a^{\dagger}_{{\bf k}1}a_{{\bf k}1}
-a^{\dagger}_{{\bf k}2}a_{{\bf k}2}]\right.\nonumber\\
&+&\left.E_{{\bf k}2}[a^{\dagger}_{{\bf k}3}a_{{\bf k}3}
-a^{\dagger}_{{\bf k}4}a_{{\bf k}4}]\right\}-\mu
\sum\limits_{{\bf k},i=1}^{4}a^{\dagger}_{{\bf k}i}a_{{\bf k}i}.
\label{A4}
\end{eqnarray}
\subsection{CE-structure}
First, we rewrite the linearized MF Hamiltonian (\ref{13}) in the
momentum space:
\end{multicols}
\widetext
\begin{eqnarray}
H^{\rm CE}_{\rm el}
=&-&\sum_{{\bf k},\alpha,\beta}
\left\{
\left[
t_{1}^{\alpha\beta}
\{a_{{\bf k}\alpha }^{\dagger}b_{{\bf k}\beta}+{\bar a}_{{\bf k}\alpha }^{\dagger}
{\bar b}_{{\bf k}\beta}\}\right.+\left.t_{2}^{\alpha\beta}
\{b_{{\bf k}\alpha }^{\dagger}{\bar a}_{{\bf k}\beta}+e^{i{\bf k}}
{\bar b}_{{\bf k}\alpha }^{\dagger}
a_{{\bf k}\beta}\}+H.c.\right]
\right.\nonumber\\
&-&(\mu+\Delta)
\{a_{{\bf k}\alpha }^{\dagger}a_{{\bf k}\alpha}+
{\bar a}_{{\bf k}\alpha }^{\dagger}{\bar a}_{{\bf k}\alpha}\}-(\mu-\Delta)
\left.\{b_{{\bf k}\alpha }^{\dagger}b_{{\bf k}\alpha}-
{\bar b}_{{\bf k}\alpha }^{\dagger}{\bar b}_{{\bf k}\alpha}\}
\right\}
\label{A5}
\end{eqnarray}
\begin{multicols}{2}
\narrowtext
where $ -\pi\leq k\leq\pi$ and the lattice constant is set to be unity
($a^{\prime}=1$ see Fig.\ref{f4}).
The above Hamiltonian can be simplified by transforming to new fermion operators
$\left\{\xi_{{\bf k}\alpha}, {\bar \xi}_{{\bf k}\alpha},
\eta_{{\bf k}\alpha}, {\bar \eta}_{{\bf k}\alpha}\right\}$,
\begin{equation}
a_{{\bf k}\alpha}=\frac{\xi_{{\bf k}\alpha}+{\bar \xi}_{{\bf k}\alpha}}{\sqrt{2}},\;\;\;
{\bar a}_{{\bf k}\alpha}=e^{i{\bf k}/2}\frac{\xi_{{\bf k}\alpha}+{\bar \xi}_{{\bf k}\alpha}}{\sqrt{2}}.
\label{A6}
\end{equation}
The operators $\left\{\eta_{\bf k}, {\bar \eta}_{\bf k}\right\}$,
are obtained from $b_{{\bf k}\alpha},{\bar b}_{{\bf k}\alpha}$
by the same transformation as in Eq.(\ref{A6}).
One finds the effective Hamiltonian of the form
$H^{\rm CE}_{\rm el}=H+{\bar H}$, where
\begin{eqnarray}
H=&-&\sum_{{\bf k}\alpha\beta}\left\{t^{\alpha\beta}_{1}
\xi_{{\bf k}\alpha}^{\dagger}\eta_{{\bf k}\beta}+
t^{\alpha\beta}_{2}
e^{i{\bf k}/2}\eta_{{\bf k}\alpha}^{\dagger}\xi_{{\bf k}\beta}+
+H.c.\right\}\nonumber\\
&-&\sum_{{\bf k}\alpha}\left\{(\mu+\Delta)
\xi_{{\bf k}\alpha }^{\dagger}\xi_{{\bf k}\alpha}+(\mu-\Delta)
\eta_{{\bf k}\alpha }^{\dagger}\eta_{{\bf k}\alpha}\right\},
\label{A7}
\\
{\bar H}&=&H\left[\xi(\eta)\rightarrow
{\bar\xi}({\bar \eta}),t^{\alpha\beta}_{2}\rightarrow -
t^{\alpha\beta}_{2}\right] \; .
\label{A8}
\end{eqnarray}
Further, we consider only the first part given by  (\ref{A7}),
generalization of the
diagonalization procedure on ${\bar H}$
is straightforward. With the help of the explicit expression of
the  hopping amplitude matrix we first transform operators $\eta_{{\bf k}1,2}$ to
${\tilde \eta}_{{\bf k}1,2}$ by the fermionic $u-v$ transformation
with
\begin{equation}
u_{\bf k}=\frac{1-e^{i{\bf k}/2}/2}{\Omega_{\bf k}},\;\;
v_{\bf k}=\frac{\sqrt{3}e^{i{\bf k}/2}/2}{\Omega_{\bf k}},\;\;
\label{A9}
\end{equation}
where $\Omega_{\bf k}=\sqrt{2-\cos{\bf k}/2}$.
In terms of the new operators the effective Hamiltonian is written as
\begin{eqnarray}
H=&-&\sum_{{\bf k}}{\cal E}_{\bf k}\{
\xi_{{\bf k}1}^{\dagger}{\tilde \eta}_{{\bf k}1}
+H.c.\}\nonumber\\
&-&\sum_{{\bf k}\alpha}\left\{(\mu+\Delta)
\xi_{{\bf k}\alpha }^{\dagger}\xi_{{\bf k}\alpha}+(\mu-\Delta)
{\tilde \eta}_{{\bf k}\alpha }^{\dagger}{\tilde \eta}_{{\bf k}\alpha}\right\}.
\label{A10}
\end{eqnarray}
where ${\cal E}_{\bf k}=t\Omega_{\bf k}$. The transformed Hamiltonian
is already diagonal in the subspace given by the
effective orbital index  $\alpha=2$.
However it mixes the
fermionic fields $\xi_{{\bf k}\alpha}$ and
$\eta_{{\bf k}\alpha}$ at $\alpha=1$.
The hybridization part of the Hamiltonian ({\ref{A10})
can be diagonalized  following  the same rout as in Eqs.
(\ref{A1},\ref{A2}). As a result we come
to the following diagonal form:
\begin{eqnarray}
H&=&\sum\limits_{{\bf k}}\left\{
{\cal E}_{{\bf k}}[\beta^{\dagger}_{{\bf k}1}\beta_{{\bf k}1}
-\beta^{\dagger}_{{\bf k}2}\beta_{{\bf k}2}]\right.\nonumber\\
&+&\left.\Delta[\beta^{
\dagger}_{{\bf k}3}\beta_{{\bf k}3}
-\beta^{\dagger}_{{\bf k}4}\beta_{{\bf k}4}]\right\}-\mu
\sum\limits_{{\bf k},i=1}^{4}\beta^{\dagger}_{{\bf k}i}\beta_{{\bf k}i},
\nonumber\\
{\bar H}&=&H\left[\beta\rightarrow
{\bar \beta},{\cal E} \rightarrow {\bar {\cal E}}\right]
\label{A11}
\end{eqnarray}
where
\begin{eqnarray}
{\cal E}_{\bf k}=\sqrt{\Delta^2+t^2(2-\cos k/2)},\nonumber\\
{\bar {\cal E}}_{\bf k}=\sqrt{\Delta^2+t^2(2+\cos k/2)},
\label{A12}
\end{eqnarray}
and $\beta({\bar \beta})_{{\bf k}3}={\tilde \eta}({\tilde{\bar \eta}})_{{\bf k}2}$, $\beta({\bar \beta})_{{\bf k}4}=\xi({\bar \xi})_{{\bf k}2}$, and
 $\beta({\bar \beta})_{{\bf k}1,2}$ are given by the linear combination
of $\xi({\bar \xi})_{{\bf k}1}$ and 
${\tilde \eta}({\tilde{\bar \eta}})_{{\bf k}1}$.

Let us now briefly comment the obtained band structure. It is given by the
bonding and antibonding bands
 $\mp{\cal E}({\bar {\cal E}})_{\bf k}$ (\ref{A12}) and nonbonding bands 
$\pm \Delta$ in between. The nondirectional orbital from the middle sites 
($d_{y^2-z^2}$ and $d_{x^2-z^2}$ orbitals on $a$ and ${\bar a}$
type sites, respectively) is completely decoupled from the other states giving rise to the nonbolding band with energy $-\Delta$ corresponding to 
$\beta({\bar \beta})_{{\bf k}4}$ in Eq.(\ref{A11}). The other nonbonding band with energy 
$\Delta$ corresponding to the states
$\beta({\bar \beta})_{{\bf k}3}$ is given by the linear combination of the 
degenerate orbitals on the corner sites. The state orthogonal 
to this nonboning state
 hybridizes with the directional orbital on the middle sites ($d_{3x^2-r^2}$ 
and $d_{3y^2-r^2}$ orbitals on $a$ and ${\bar a}$ type sites, respectively) 
leading to the bonding and antibonding bands. 
As follows from the above discussion, the particular geometry of 
the zig-zag structure not only leads to the opening of ``dimerization''-like 
gap in the spectrum, but also removes the orbital degeneracy on the middle 
sites. The energy of the directional orbital is lowered 
due to the  hybridization and hence  delocalization, 
while its orthogonal orbital remains local.

}\end{multicols}
\end{document}